\documentclass[article,12pt,nofootinbib]{revtex4}
\usepackage[paper=letterpaper,margin=1.5in]{geometry}
\usepackage{epsfig,color}
\usepackage{graphicx}
\usepackage{amssymb,amsmath}
\usepackage{time}
\usepackage[varg]{txfonts}
\usepackage{hyperref}
\hypersetup{
  colorlinks,
  citecolor=green,
  linkcolor=blue}
\newcommand{\be}{\begin{equation}}
\newcommand{\ee}{\end{equation}}
\newcommand{\bc}{\begin{center}}
\newcommand{\ec}{\end{center}}

\renewcommand{\vec}[1]{\textnormal{\boldmath$#1$}}
\begin{document}

\bibliographystyle{revtex}


\vspace{.4cm}

\title
{Experimental validation of collective effects modeling at injector section of x-ray free-electron laser
}
\author{Igor  Zagorodnov,  Sergey Tomin, Ye Chen, Frank Brinker}
\affiliation{Deutsches Elektronen-Synchrotron, Notkestrasse 85,
22603 Hamburg, Germany}
\date{\today}

\vspace{.4cm} 
\begin{abstract} 
We consider the collective  beam dynamics at the injector section of the European XFEL. The results of the measurements for the longitudinal phase space (LPS) of the electron beam are compared with those obtained from numerical modeling. A new approach is proposed for the analysis of the self-field effects in the LPS measurements. It allows to determine accurately the synchronous RF phase in an accelerating module and to subtract properly the RF curvature imprinted in the LPS. A further incorporation with the simulation made it possible to separate the collective effects originating from different sources and thus to quantify, individually, the impact of these effects on the beam dynamics. This includes the space-charge dominated beam dynamics in the RF gun as well as the collective effects dominated by the wake fields after the gun up to the end of the injector section.  A new analytical model is also proposed for the short-range wake function of a finite chain of RF cavities and verified by both numerical analysis and experimental results.  It is shown that the physical models used in the simulation of the beam dynamics after the gun allow to reduce the absolute error in the modeling of the correlated energy chirp by order of magnitude in comparison with the case when the collective effects after the gun  are neglected.

\end{abstract}

\maketitle

\section{Introduction}\label{sec:1}
 
 The numerical modeling of modern free electron lasers remains a challenge for the computational physics. The different scales of the physical processes do not allow to use self-consistent simulations of the particle motion with direct solution of the Maxwell's equations along the whole machine. Several approximate models like "space charge field", "wake field", "coherent synchrotron radiation", "paraxial field approximation"  are developed and used in numerical codes to estimate the properties of the electron bunch at the end of the electron accelerator. The models are approximate and only experiments could confirm that they describe the collective beam dynamics accurately. Or the experiment can at least help us estimate the errors of the modeling. 

However,  direct comparison between simulation and measurement is very difficult as many technical parameters of the machine are known only approximately or with insufficient accuracy. For example, a small error in the definition of the radio frequency (RF) parameters of the booster (the first RF module after RF gun)  changes the bunch properties at the end of the linac enormously~\cite{Zag19}. The statistical errors usually can be excluded easily from the experiment, but the systematic errors due to the calibration or drifting are difficult to identify and to remove. Hence methods of the analysis which allow to remove the systematic errors are of great importance.

The accelerator beam dynamics of the European X-ray Free Electron Laser (XFEL) ~\cite{EXFEL}  has been discussed recently in paper ~\cite{Zag19}. The authors have described the physical models used and have presented results of numerical modeling. In this paper we continue experimental and numerical studies of the beam dynamics and concentrate our attention on the injector section of the facility: from the radio frequency gun up to the dogleg section before the first bunch compressor. Due to a relatively low energy of the electron beam at this section the collective effects impact considerably the beam dynamics and an accurate verification of the physical models applied in the simulations is highly desired. 

In this paper we present a new approach to the analysis of the self-field effects in the measurements of the longitudinal phase space (LPS). It allows to identify the  synchronous RF phase in an accelerating module with a high accuracy and to subtract properly the RF curvature imprinted in the longitudinal phase space. In combination with beam dynamics simulations it allows to  separate the impact of the collective effects originating from different sources: the space-charge dominated beam dynamics in the RF gun and the collective effects dominated by the wake-fields after the gun up to the end of the injector section.   

{The measurement of LPS with a transverse deflecting structure (TDS) is a well established procedure~\cite{TDS}. The main challenge is the analysis of the measurements and the reproducibility  in the numerical modeling. The energy chirp or the curvature introduced by RF is considerably larger than that introduced by collective effects. Hence we need an accurate procedure to define the RF parameters. The standard procedure (scan of the RF phase and measurement of the beam energy in the dispersion section) has a considerable error (approximately 0.1 degree) in definition of RF phase. Additionally, the energy measurement used for the calibration of RF voltage has a considerable error (approximately 1\%) as well. In order to overcome these difficulties and to substract the RF curvature with high accuracy in the post-processing we suggest a new method. It is based on the analysis of TDS measurements with the hypothesis that the collective effects are independent from the RF phase if the beam energy is kept approximately constant. To our knowledge  this accurate and reliable method of the extraction of the RF curtvature was not described in the literature so far. }

The results of the measurements {and the analysis} are compared with those of the numerical modeling.  In order to take into account properly the wake field effects in the accelerator modules we have developed a new model of the wake functions and have done numerical studies to find out the coefficients of the model for the different RF modules available in the injector section.  The new model is an extension of the model suggested earlier in~\cite{Zag04}. The old model was not able to describe accurately the wake behavior required for the long bunches at the injector section. Hence in this paper a new model with additional terms is introduced. It is shown for three different geometries  that the new model is able to reproduce  accurately the numerical results obtained by direct solution of the Maxwell's equations. We have found that after the RF gun the wake fields have a stronger impact on the bunch than the longitudinal space charge forces. 

The comparison of the theoretical collective effects with the measured ones in XFEL facilities has been done, for example, in~\cite{Zag19, Dohlus, diMitri}. However, the authors of these and other publications usually compare only 
the integrated parameters (energy loss, rms energy spread, transverse kick) with the measured ones. In this paper we carry out a time-resolved comparison of theoretical and measured collective effects inside of the electron bunch. Such kind of study requires much more accurate and advanced measurements and allows to prove the theoretical {models used in the simulations. We show that the physical models used in the simulation of the beam dynamics after the gun allow to reduce the absolute error in the modeling of the correlated energy chirp by order of magnitude in comparison with the case when the collective effects after the gun  are neglected.}

We start in Section~\ref{sec2} from the layout of the  injector section used in the measurements and the simulations. Then, in Section~\ref{sec3} we describe the approach to the measurements and to their analysis. In Section~\ref{sec4} we describe a model of wake field functions for the finite chain of RF modules and present the calculations to obtain the coefficients of the model for the structures available in the injector section. These results are used in the beam dynamics simulations described in Sections~\ref{sec4}, \ref{sec5}. In Section~\ref{sec5} we compare the numerical results with the measured ones and confirm the correctness of the physical models used in the simulations.  
  
\section{Layout of the injector section and the setup of the measurements}\label{sec2}

\begin{figure}[htbp]
	\centering
	\includegraphics*[height=35mm]{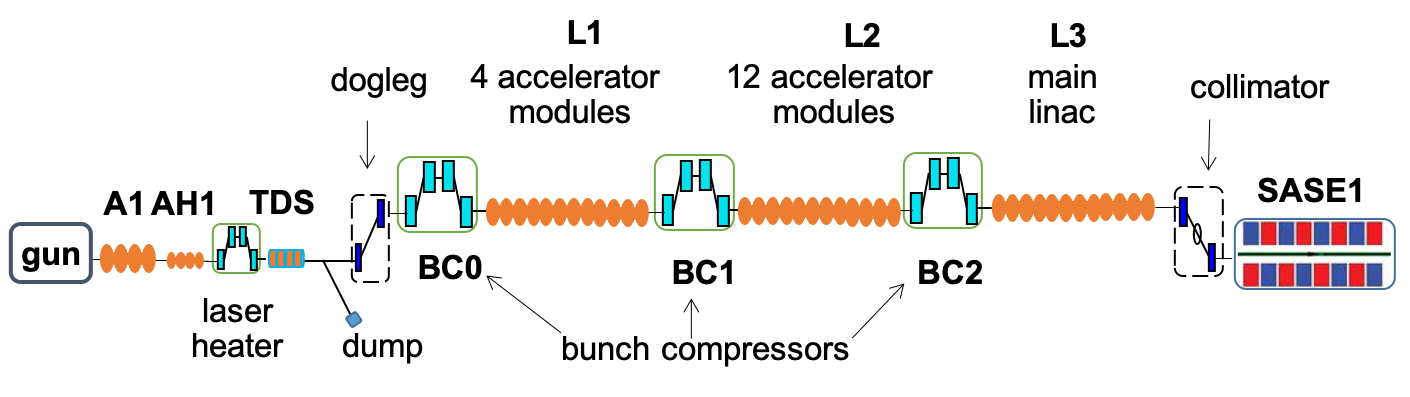}
	\caption{The layout of the European XFEL accelerator  with SASE1 branch.}\label{Layout}
\end{figure}

The European XFEL produces the hard X-rays down to sub-angstrom wavelength. It consists of a 17.5 GeV linear accelerator (linac) and several undulator lines for SASE radiation, namely, SASE1, SASE2 and SASE3. The layout of the facility with the SASE1 branch is shown in Fig.~\ref{Layout}. 

\begin{figure}[htbp]
	\centering
	\includegraphics*[height=30mm]{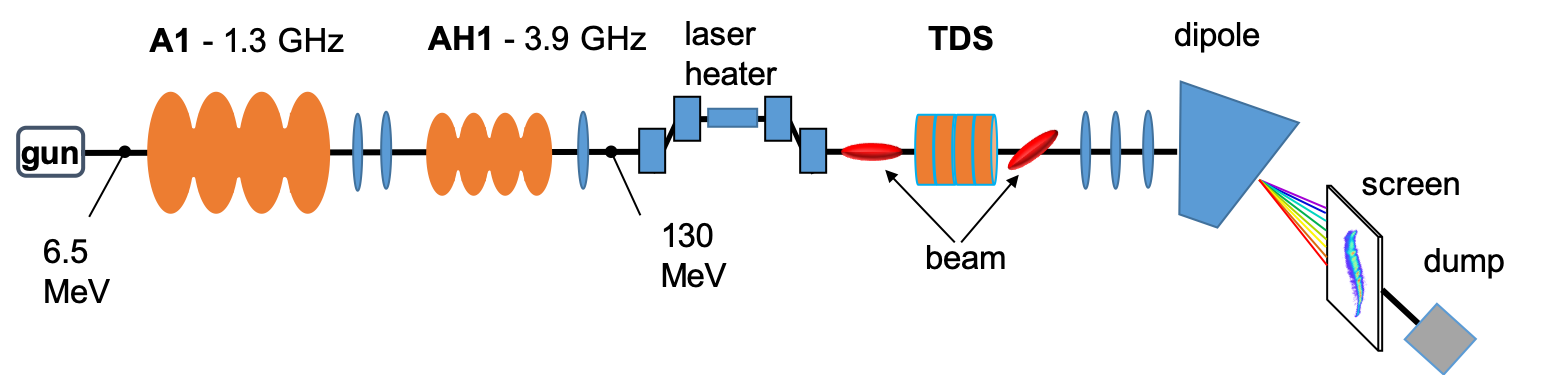}
	\caption{The setup of the experiment in the injector section.}\label{Setup}
\end{figure}

The injector section starts from the gun and ends before the bunch compressor BC0. In our studies we use only the part of the injector section shown at Fig.~\ref{Setup}. The electron bunch travels a distance of 40 meters from the cathode of the RF gun up to the OTR screen in the injector dumpline. In conventional operation of the facility the TESLA cryomodule A1 accelerates the beam up to 150 MeV and then the bunch is decelerated in the high harmonic cryomodule AH1 to the energy of 130 MeV. With this energy the bunch travels up to the bunch compressor BC0 for the compression to a higher peak current.

In the experiment presented in this paper we switched off the high harmonic module AH1 and the laser heater. During the scans of different parameters we kept the beam energy near to 130 MeV already after the first cryomodule A1. The bunch was transversely kicked in the TDS and analyzed on the OTR screen after the dipole magnet in the dumpline. In order to calibrate the time and energy axes of the longitudinal phase space we have proceeded in the usual way by using the phase of the TDS and the voltage of cryomodule A1.

\begin{table}[htbp]
	\centering
	\caption{Parameters of the measurement setup.}
	\label{TableMS}
	\label{Table_SP}
	\begin{tabular}{lcccll}
		\hline\hline
		{\bf parameter}& 	{\bf Units}& 	{\bf Value}\\
		\hline
		OTR resolution, $\sigma_R$  &$\mu$m&20\\
		Normalized emittance, $\epsilon_n$  &$\mu$m&0.6\\
		Optical  $\beta$-parameters at OTR , $(\beta_x,\beta_y)$  &m&(0.6,15)\\
		Dispersion,  $D$  &m&0.59\\
		Optical $\beta$-parameter at TDS,  $\beta_y^0$  &m&3\\
		Optical $\alpha$-parameter at TDS,  $\alpha_y^0$  & &1.47\\
		Wave number of TDS,  $k$  & 1/m&58.7\\
		Voltage of TDS,  $V$  & MV&0.84\\
		Length of TDS,  $L$  & m&0.7\\
		Streak of TDS,  $S$  & &1.8\\
		Reference energy,  $E_0$  & MeV&130\\
		\hline\hline
	\end{tabular}
\end{table}

{In order to  estimate the resolutions of the measurements in time $R_t$ and in energy $R_E$ we use the following equations
\begin{align}\label{Eq_res}
	R_t&=\frac{1}{S c}\sqrt{\sigma_R^2+\sigma_{y}^2},\quad   \sigma_{y}^2=\frac{\beta_y \epsilon_n}{\gamma_0},\\
	R_E&=\frac{E}{D}\sqrt{\sigma_R^2+\sigma_x^2+(D K \sigma_I)^2}, \quad K=\frac{e k V}{E_0},\label{Eq_K}\\ \sigma_{x}^2&=\frac{\beta_x \epsilon_n}{\gamma_0},\quad  \sigma_{I}^2=\frac{(\beta_y^0+0.25 L^2 \gamma_y^0-L\alpha_y^0) \epsilon_n}{\gamma_0},
\end{align}
where $e$ is the absolute value of electron charge, $c$ is the velocity of light in vacuum, $\gamma_0$ is the relative relativistic energy,  $\gamma_y^0$ is the value of the optical function at the middle of TDS and the meanings and the values of other  parameters are listed in Table~ \ref{TableMS}. With these values we estimate that in the measurements presented in the following sections  the time resolution $R_t$ is equal to 0.35 ps and the energy resolution $R_E$ is equal to 10 keV.}

\section{Measurements and their analysis}\label{sec3}

{The longitudinal component of the electric field on the axis of the gun cavity can be represented as
\begin{align}\label{Eq_gun}
E_z(z,t)=E_z(z)\sin(\omega t +\theta),
\end{align}
where $\omega$ is the angular frequency, $\theta$ is the phase of the wave. For electrons starting at time $t=0$, the longitudinal field has the highest  value at phase of $90^o$. At phase $\theta_0=0^o$ the field on cathode changes the sign and we will call phase $\theta_0$ "zero-crossing" phase. 

In the following we present the phase $\theta$  in the form
\begin{align}\label{Eq_gun_phase}
	\theta=\theta_0 +\theta_{\text{MM}}+\Delta\theta,
\end{align}
where $\theta_{\text{MM}}$ is the phase of maximal mean momentum of the electron beam.  The "zero-crossing" phase $\theta_0$ was defined experimentally in the same way as described  at Section III.C of~\cite{Chen}.}

At the first step we have done a set of measurements to define a working point of the gun at a bunch charge $Q=250$ pC. We have found that the gun phase of maximal mean momentum of the electron beam $\theta_{MM}$ is 44 degrees and the momentum of the beam after the gun is equal to 6.45 MeV/c. 

The change of the longitudinal momentum in cryomodule A1 can be approximated by expression
\begin{align}\label{Eq_A1}
	\Delta p_z(\tau)=\frac{e V}{c}\cos(\omega \tau + \phi),\quad \omega=2 \pi f,
\end{align}
where $\tau$ is the relative time offset to the time of beam center of mass, $e$ is the electron charge, $c$ is the vacuum light velocity, $V$ and $\phi$ are the voltage and the synchronous phase of the cryomodule A1, respectively, and $f=1.3$ GHz.

The best beam emittance is obtained not in the gun phase of maximal electron beam momentum but at a phase shifted by 2-3 degree. Hence we set the gun phase  shift  $\Delta\theta$ to -2 degree. Additionally we have done a scan of the gun solenoid strength {to set the  transverse projected emittance to the minimum value of 0.6 $\mu$m.} The "emittance" means in this paper the root mean square (rms) normalized emittance. For example, in horizontal plane the projected normalized rms emittance is calculated from the total particle distribution by the formula $\epsilon_x^{proj}=(m_{0} c)^{-1}\sqrt{\left< x^2\right>\left<p_x^2\right> - \left< x p_x\right>^2}$, where $p_x$ is the particle momentum and symbol $\left< \right>$ defines the second central moment of the particle distribution.

At this state of the gun we have done a scan of the phase of cryomodule A1 to define the RF phase of maximal mean momentum of the electron beam. The longitudinal electric field on the axis of cryomodule A1 has the same form as given by Eq.~(\ref{Eq_gun}) with RF phase $\theta_{A1}$. The RF phase of the cryomodule A1 was set to this phase $\theta_{A1}$. Note that, the exact value of the RF phase is not important for our discussion. We use only the fact that for the RF phase $\theta_{A1}$ (so called "on-crest" phase) the synchronous phase $\phi$ is zero for the bunch charge $ Q=250$ pC and the gun phase offset $\Delta\theta=-2^o$.  

\begin{figure}[htbp]
	\centering
	\includegraphics*[height=55mm]{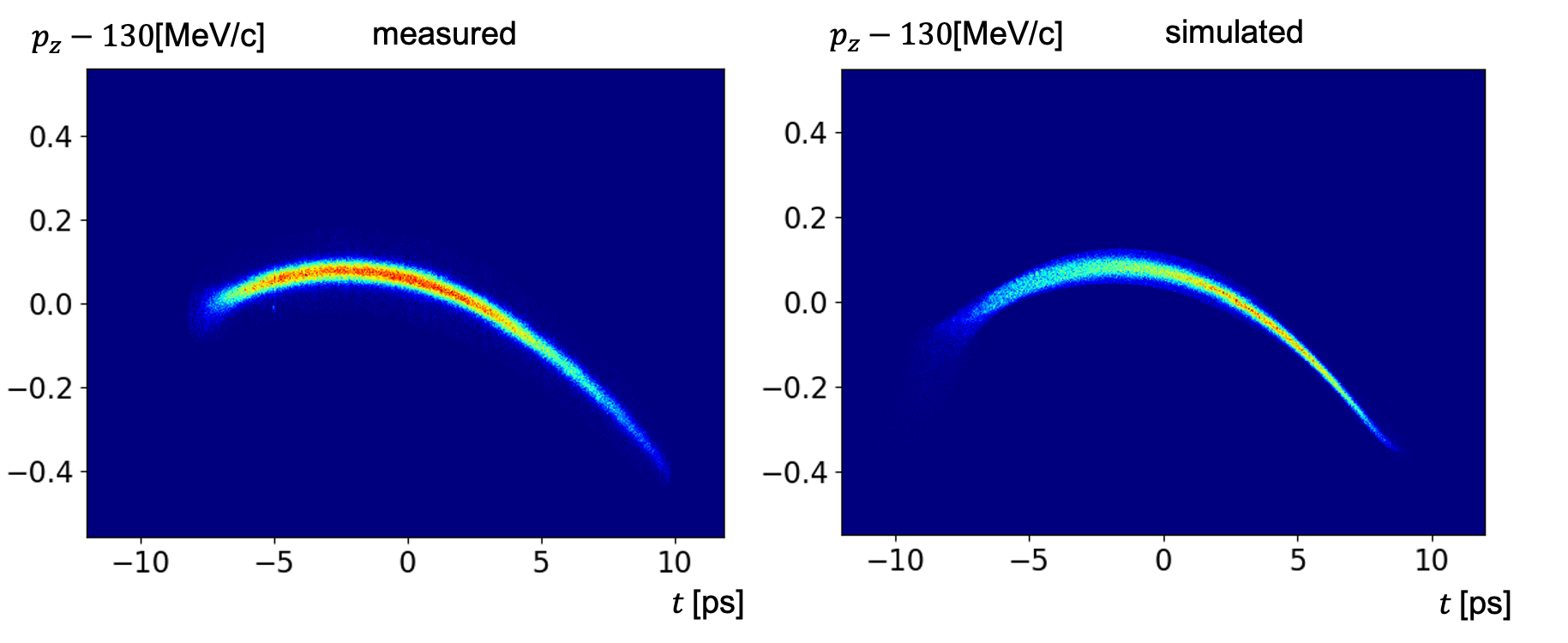}
	\caption{Comparison of the longitudinal phase space from  measurement (left) and from simulation (right) for the bunch charge $Q=250$ pC , the gun phase offset $\Delta\theta=-2^0$ and the synchronous phase of cryomodule A1 $\phi=0$. }\label{Fig003}
\end{figure}

{The left  image in Fig.~\ref{Fig003} presents the phase space measured with TDS at the gun phase offset $\Delta\theta=-2^o$ and the synchronous phase $\phi=0^o$. The right image in  Fig.~\ref{Fig003} is obtained from beam dynamics simulations which include collective effects, impact of TDS and simulation of the OTR screen in the dump. } We see an asymmetric curvature of the phase space which does not fit the cosinus function shape of Eq.~(\ref{Eq_A1}). It means that the electron beam has a considerable energy chirp induced by the RF in the gun and the collective forces in the gun and further downstream. Hence the synchronous phase of the maximal mean momentum is different from the synchronous phase of the smallest energy spread. The smallest energy spread will be reached at a phase which makes the LPS shape symmetric (the head and the tail of the bunch have the same energy). The A1 phase of minimal energy spread is usually different from the phase of maximal beam momentum by approximately 1 degree.

In order to compare the measurement with the simulation and to verify the physical models of different collective effects we need to make not only one measurement but a whole set of scans for different parameters. In the experiment we have done measurements for three different bunch charges (250 pC, 500 pC, 750 pC) and for the gun phases offset $\Delta\theta$ from -8 to 4 degrees. The scan of the gun phase is necessary for two reasons: (1) to define the RF phase $\theta_{\text{MM}}$ of maximal mean momentum and (2) to check that the energy chirp from the gun is correctly reproduced in the simulations.

If we change the bunch charge $Q$ or the gun phase offset $\Delta\theta$ then the synchronous phase $\phi$ of the cryomodule A1 will change as well. We need to know this phase with high accuracy to be able to compare the measurement data with the simulation results. Defining of the synchronous phase $\phi$ usually is carried out experimentally by looking for the RF phase with the maximal electron beam momentum. Such approach requires a considerable amount of operational time of the facility and the accuracy usually is not better than 0.1 degree. 

During the experiment we have decided to use another method and to extract the synchronous phase $\phi$ from the measured data in the post-processing. We have found that the method presented below allows saving the facility operation time and to obtain, with a higher accuracy, the synchronous phase $\phi$ than the conventional method, where an experimental searching approach needs to be carried out, iteratively, for each new setup (a new bunch charge or a new gun phase), to find the RF phase of the maximum beam momentum.

\begin{figure}[htbp]
	\centering
	\includegraphics*[height=55mm]{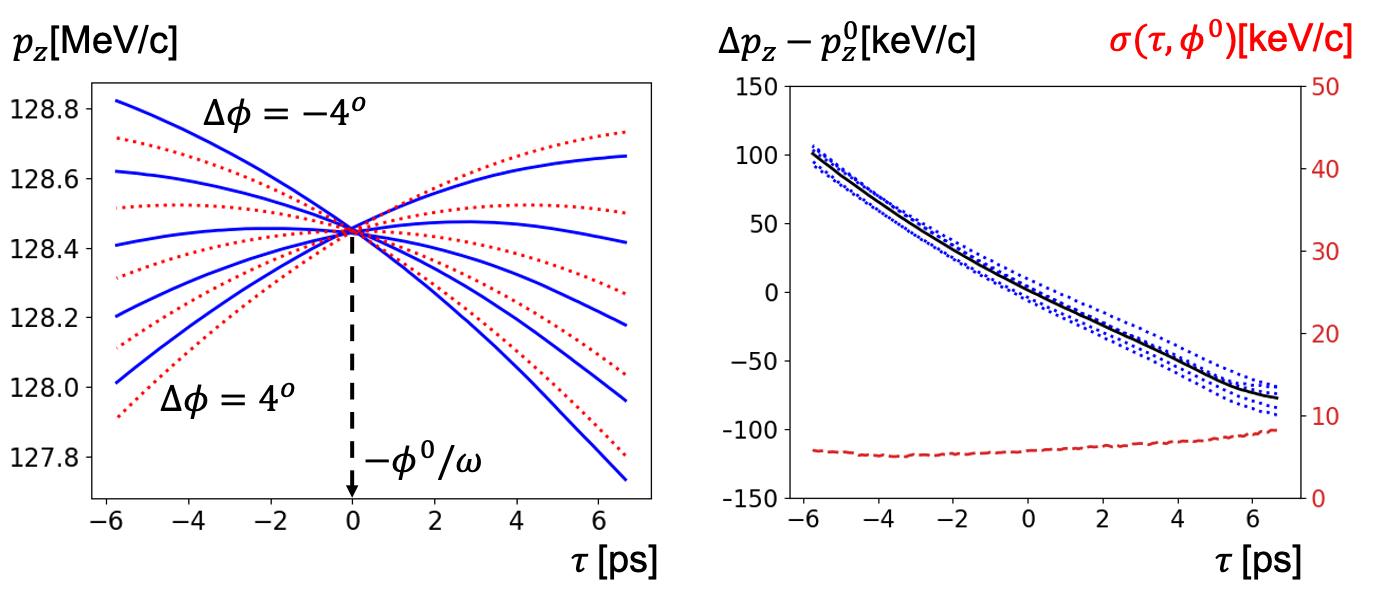}
	\caption{The left plot presents the measurement of the mean slice momentum for the bunch charge $Q= 250$ pC and the gun phase offset  $\Delta\theta=-2^o$. The right plot shows the residual mean slice momentum after a subtraction of the RF curvature (the dotted red curves in the left plot). The solid black line in the right plot presents the averaging of the dotted curves (see Eq.~(\ref{Eq_dpz})). The dashed red line presents the standard deviation as defined by Eq.~(\ref{Eq_sigma}).}\label{Fig004}
\end{figure}

At each working point of interest (the bunch charge and the gun phase) we change the RF phase of cryomodule A1 by $\pm 4^o$ and save the longitudinal phase space (LPS) by recording the beam image on the OTR screen in the injector dumpline. If we suggest that the residual LPS after a subtraction of the RF curvature of cryomodule A1 is independent from the RF phase of A1, then from a simple analysis in post processing of the measured data we are able to identify the synchronous phase $\phi$ accurately and to subtract the RF curvature of the cryomodule properly from the LPS.  {Note, that another  hypothesis which we use is the suggestion that for the relativistic beam the change in the syncronous phase $\phi$  is equal to the change in the RF phase $\theta_{A1}$. In the following we will use for  both offsets in the phase  the same symbol $\Delta\phi$.

In the following we will show only a mean position of the slices in the longitudinal phase space
\begin{align}\label{}
	p_z(\tau)&= \frac{1}{\rho(\tau)}\int p_z f(\tau,p_z)d p_z,\\ 
	\rho(\tau)&=\int f(\tau,p_z)d p_z,\quad
	\iint f(\tau,p_z)d \tau d p_z=1,
\end{align}
where $f(\tau,p_z)$ is the phase space density of the electron beam in the longitudinal phase space (see, for example, Fig.~\ref{Fig003}).

On the left plot of Fig.~\ref{Fig004}, we show by solid blue lines the measured mean slice momentum	$p_z(\tau)$ at different RF phases of cryomodule A1. The measured curves can be presented in the form
\begin{align}\label{Eq_RF}
p_z(\tau,\Delta \phi)=  \frac{e V(\Delta \phi)}{c}\cos(\omega \tau +\Delta \phi+\phi^0)+\Delta p_z(\tau, \Delta \phi, \phi^0).
\end{align}
On the left hand side the function $p_z(\tau,\Delta \phi)$ is the curve measured in the experiment for the RF phase $\theta_{A1}+\Delta \phi$ and obtained by processing of the data from the OTR screen and from the energy monitor. 

On the right hand side of Eq.~(\ref{Eq_RF}) we introduce a free parameters  $\phi^0$ to be yet defined. The parameter $\phi^0$ is the synchronous phase of cryomodule A1 when the RF phase of the module $\theta_{A1}$ has the same value as in the measurement presented in Fig.~\ref{Fig003} ($\Delta \phi=0$). The value of voltage $V(\Delta \phi)$ was adjusted during the measurement analytically $ V(\Delta \phi)=V(0)/\cos(\Delta \phi)$ to keep the beam energy near to constant value. Finally, the last term in the expression $\Delta p_z(\tau, \Delta \phi,  \phi^0)$ describes the residual mean slice momentum when the RF curvature of cryomodule A1 is subtracted. This term is created by the beam dynamics in the RF gun and by the collective effects after the gun. This term depends only weakly on the RF phase of A1 if the final energy of the beam is kept approximately constant.

In order to find the parameter $\phi^0$, which provides the residual term $\Delta p_z(\tau, \Delta \phi, \phi^0)$  independent from the phase shift $\Delta \phi$ we proceed as follows. We calculate the mean and the rms residual functions
\begin{align}\label{}
m(\tau,\phi^0)&= \frac{1}{N}\sum_ {\Delta \phi}\Delta p_z(\tau, \Delta \phi,  \phi^0),\\
\sigma(\tau,\phi^0)&= \bigg(\frac{1}{N-1}\sum_ {\Delta \phi}\big[\Delta p_z(\tau, \Delta \phi,  \phi^0)-m(\tau,\phi^0)\big]^2\bigg)^{1/2},\label{Eq_sigma}
\end{align}
where $N$ is the number of different phase offsets $\Delta \phi_{A1}$ used.

The value $\phi^0$ is defined as those that minimize the averaged value of the rms residual function :
 \begin{align}\label{}
\phi^0: \min\limits_{\phi}  \int \sigma(\tau,\phi) d\tau.
 \end{align}
For the phase  $\phi^0$ we can define the residual mean slice momentum $\Delta p_z( \tau)$ and the residual mean momentum $p_z^0$:
\begin{align}\label{Eq_dpz}
	\Delta p_z( \tau)= m( \tau,  \phi^0),\quad p_z^0=\int \Delta p_z( \tau)\rho( \tau) d  \tau . 
\end{align}

The parameter $p_z^0$ is a sum of initial momentum of the beam, the energy loss due to the wake fields and the errors in the energy measurement and in the RF calibration.

For the bunch charge $Q=250$ pC and the gun phase offset $\Delta\theta=-2^o$  we have found $p_z^0=5.93$ MeV/c, $\phi^0=-0.01^o$. The value of $p_z^0$ is different from the measured beam momentum after the gun, 6.45 MeV/c. One of the reasons is the energy loss due to wake fields. From the model of wake fields presented in Section~\ref{subsec4.2} we estimate that the energy loss due to the scattered fields for the bunch charge of 250 pC is equal only to 54 keV. Hence the main sources of this difference are the errors in the energy measurements  and the error in the calibration of voltage of cryomodule A1. The value of the synchronous phase $\phi^0$ is very close to zero as expected.

 The red dotted curves on the left plot of Fig.~\ref{Fig004} show the curvature induced in the longitudinal phase space of the electron beam by RF field of cryomodule A1 alone (the sum of the first term in Eq.~(\ref{Eq_RF}) and $p_z^0$). If we subtract the values of these curves from the measured ones (shown by solid blue lines) then we obtain the residual curves $\Delta p_z( \tau, \Delta \phi, \phi^0)-p_z^0$ for $\Delta \phi=-4^o,-2^o,0^o,2^o,4^o$, shown on the right plot by blue dotted lines. As expected they depend only weakly on the phase of cryomodule A1. Then, we can calculate the residual mean slice momentum $\Delta p_z(\tau)-p_z^0$ , represented by the averaged curve (solid black line). The dashed red line in the right plot presents the standard deviation $\sigma(\tau,\phi^0)$ as defined by Eq.~(\ref{Eq_sigma}).}

\begin{figure}[htbp]
	\centering
	\includegraphics*[height=55mm]{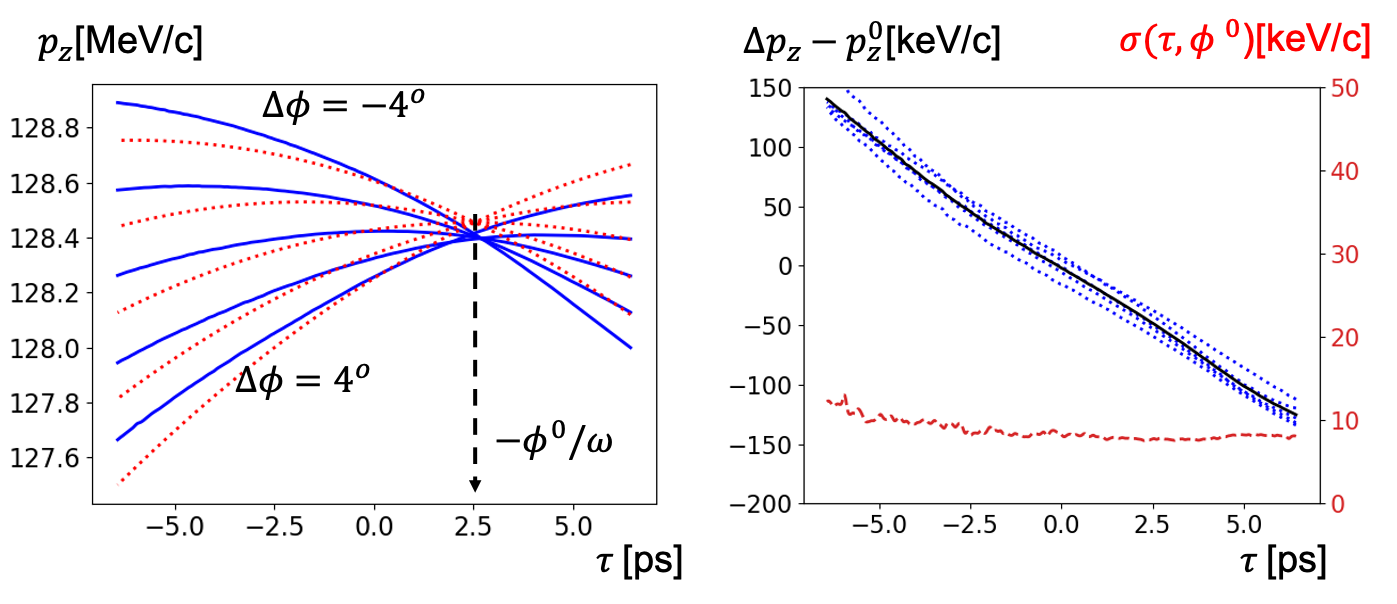}
	\caption{The left plot presents the measurement of the mean slice momentum (blue solid curves) for the bunch charge $Q= 250$ pC and the gun phase offset $\Delta\theta=4^o$. The right plot shows the residual mean slice momentum after a subtraction of the RF curvature (the dotted red curves in the left plot). The solid black line in the right plot presents the averaging of the dotted curves (see Eq.~(\ref{Eq_dpz})).The dashed red line presents the standard deviation as defined by Eq.~(\ref{Eq_sigma}).}\label{Fig005}
\end{figure}

\begin{table}[htbp]
	\centering
	\caption{The synchronous phase $\phi^0$ in degree.}
	\label{Table_A1}
	\begin{tabular}{lccccc}
		\hline\hline
		&\boldmath $-8^o$ &\boldmath $-5^o$ &
		\boldmath $-2^o$ &\boldmath $1^o$ &
		\boldmath $4^o$\\
		\hline
		{\bf 250 pC}	&1.77&	0.82 &-0.01&-0.68&-1.19\\
		{\bf 500 pC}	&2.17&	1.25&0.41&-0.21&	-0.71\\
		{\bf 750 pC}	&2.38&	1.50 &0.74&0.21&	-0.27\\
		\hline\hline
	\end{tabular}
\end{table}

In Fig.~\ref{Fig005} we show another data set for the gun phase offset $\Delta\theta=4^o$ and the same bunch charge of 250 pC. The value of the synchronous phase  $\phi^0$ is no longer zero but $-1.19^o$. The synchronous phases determined for other measurements are summarized in Table~\ref{Table_A1}.

\section{Modelling of the beam dynamics}\label{sec4}

\subsection{Modelling of the RF gun}\label{subsec4.1}
The electron bunch is produced by a shaped laser pulse in the RF gun. The laser pulse has a temporally Gaussian shape and an rms duration of about 3 ps. The parameters used in the gun simulations are listed in Table~\ref{TableInj}.

\begin{table}[htbp]
	\centering
	\caption{Injector parameters.}
	\label{TableInj}
	\begin{tabular}{lcccll}
			\hline\hline
		{\bf subsection}& 	{\bf parameter}&	\\
			\hline
	   {\bf laser} &rms length, ps& 2.7 \\
	                   &width, mm&  1-1.5\\	                   
	   {\bf RF cavity} &frequency, GHz& 1.3 \\
	                   &maximal field on cathode, MV/m& 56.3 \\	                   
	                   & phase, degree& 35-47\\      
	     {\bf solenoid} &Magnetic field, T& 0.216 \\
	     	\hline\hline
	\end{tabular}
\end{table}

The simulations are done with code ASTRA~\cite{ASTRA}. In the gun simulations we have followed the route used experimentally in setting up the gun. The scan of the RF gun phase for the bunch charge of 250 pC resulted in the maximal mean momentum phase $\theta_{\text{MM}}$ of $43^o$, which agrees reasonably with the phase defined in the experiment. {The "zero-crossing" phase $\theta_0$ (see Eq.~(\ref{Eq_gun_phase})) is defined for a reference particle at the center of mass of the beam distribution (each particle has its own birth time when it starts from the cathode and the reference particle has started at an average time through all particles).} The strength of the solenoid field was chosen the same way as in the experiment to produce a transverse emittance of 0.6 $\mu$m after cryomodule A1. {The solenoid strength in the experiment was 0.203 T. We think that the difference in the solenoid strength used in the experiment and the simulations is due to calibration errors of the solenoid strength in the machine}.

In our simulations we use $4\times10^5$ macro-particles. As for calculating the slice parameters we have used $5\times10^3$ particles per slice. These calculations are done accordingly for the same set of bunch charges as in the experiment: 250 pC, 500 pC and 750 pC. For each charge set point, the phase $\theta_{\text{MM}}$ (see Eq.~(\ref{Eq_gun})) is set as $43^o$ and 5 different gun phase offsets $\Delta\theta$ are considered: $-8^o, -5^o, -2^o, 1^o, 4^o$ . The electron bunches are tracked through the RF gun in presence of space charge forces till the entrance of cryomodule A1. The obtained particle distributions are then used for further particle tracking as well as the analysis as described in the next sections. 

\subsection{Wake function of a finite chain of accelerating cavities}\label{subsec4.2}

In order to take into account the impact of the electromagnetic fields scattered by the vacuum chamber we use wake field formalism. The main sources of the wake fields are RF cavities. In this paper we approximate the RF cavities by rotationally symmetric geometries neglecting RF couplers and higher order modes absorbers. In the case of the rotational symmetry the longitudinal wake function near the axis can be written in the form~\cite{Zotter}
\begin{align}\label{Eq_wz}
w_z(\vec{r},s)=w_z(s)+w'_{\perp}(s)\Big(x_0 x+ y_0 y\Big)+O(\Delta \vec{r}^3), \\
 w_z(s)\equiv w_z(\vec{0},s),\qquad w'_{\perp}(s)\equiv \frac{\partial^2 w_z}{\partial x\partial x_0}(\vec{0},s), 
\end{align}
where we have incorporated in one vector the transverse coordinates of the source and the witness particles, $\vec{r}=(x_0,y_0,x,y)^T$ and $s$ is a distance between these particles in the longitudinal coordinate $s=z_0-z$. 
The transverse wake function can be found through Panofsky-Wentzel theorem and has the form:
\begin{align}
\vec{w}_{\perp}(\vec{r},s)=w_{\perp}(s)\Big(x_0 \vec{e}_x+ y_0  \vec{e}_y\Big)+O(\Delta \vec{r}^2),
\end{align}
where
\begin{align}\label{Eq_ws}
w_{\perp}(s)=\int_{-\infty}^{s}w'_{\perp}(s')ds'. 
\end{align}
\begin{figure}[htbp]
	\centering
	\includegraphics*[height=50mm]{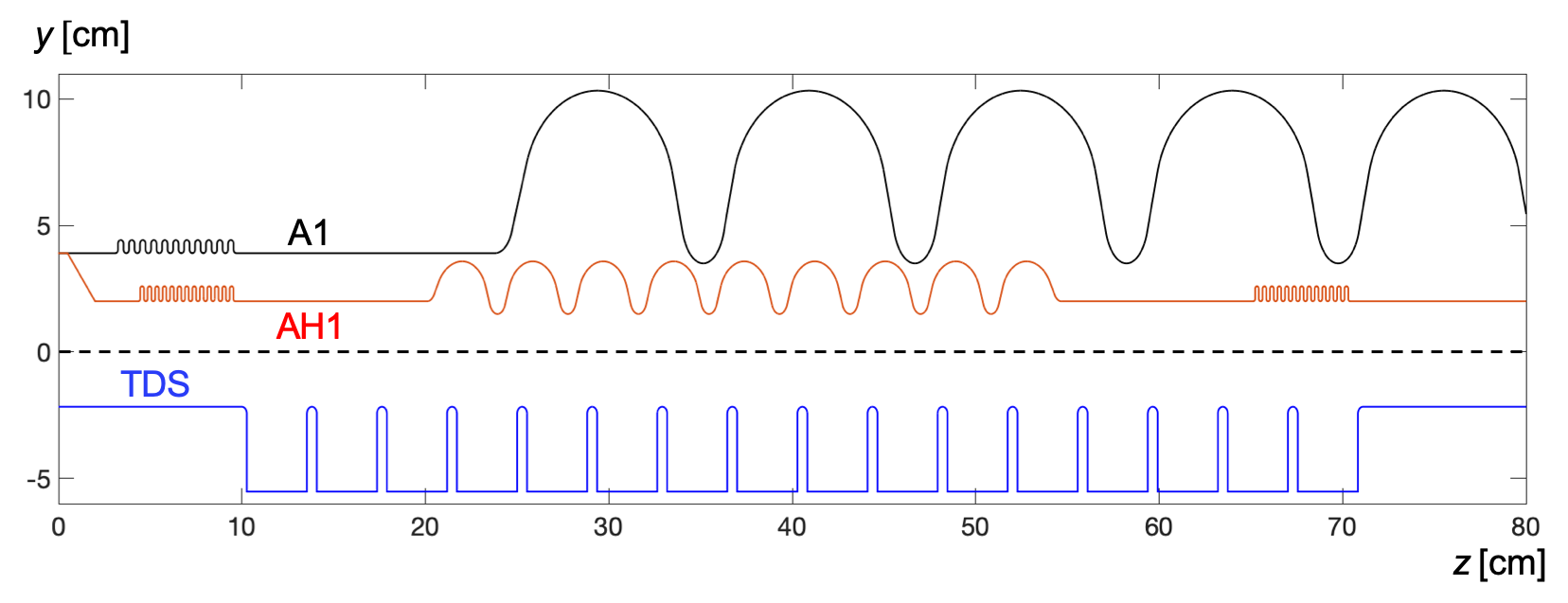}
	\caption{The geometry of RF modules installed in the injector section of the European XFEL.}\label{Fig006}
\end{figure}

\begin{figure}[htbp]
	\centering
	\includegraphics*[height=70mm]{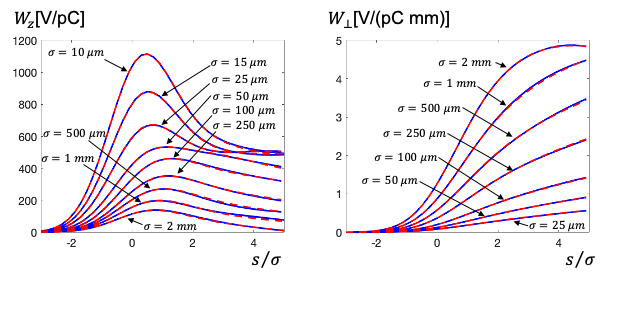}
	\caption{The longitudinal and the transverse wake potentials of the Gaussian bunches in the high harmonic cryomodule AH1. The bunch length is varied between 10 $\mu$m and 2 mm. The numerical results from code ECHO are shown by solid blue curves. The analytical approximations are shown by dashed red curves.}\label{Fig007}
\end{figure}

\begin{figure}[htbp]
	\centering
	\includegraphics*[height=60mm]{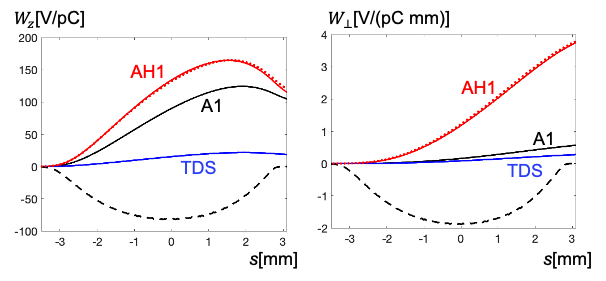}
	\caption{The longitudinal and the transverse wake potentials of the bunch with the bunch charge 500 pC at the gun phase offset of -$2^o$. The dashed line shows the bunch shape in arbitrary units. The solid curves present the wake potentials obtained from the analytical representations. The dots present the wakes for AH1 obtained by direct numerical solution of the Maxwell's equations in ECHO.}\label{Fig008}
\end{figure}

An approximate analytical forms of functions $w_{z}(s)$, $w_{\perp}(s)$ for the TESLA cryomodule (A1), the high harmonic module (AH1) and the transverse deflecting cavity (TDS) have been already found in~\cite{Novo, Zag03, Zag04} for the setups and the geometries used at FLASH.  However in~\cite{Novo, Zag03} the authors have been looking for the wake function of an infinite chain of TESLA cryomodules, where each cryomodule consists of 8 TESLA cavities connected by pipes with bellows. This infinite periodic approximation is accurate for the wakes of TESLA cryomodules in the linacs L1, L2, L3, but in the injector section the cryomodule A1 is an isolated structure and it has to be considered in the manner used in ~\cite{Zag04} for the high harmonic module and the TDS structure. At the same time the layout of the high harmonic module ~\cite{AH1} in the European XFEL is different from that described and used in~\cite{Zag03}: the high harmonic module includes not 4, but 8 cavities, the in-going step transition is replaced by a taper, the out-going step transition is removed completely. The TDS structure has a little different geometry of cells and is composed only from 16 cells~\cite{TDS1} . 

The rotationally symmetric approximation of these RF structures is presented in Fig.~\ref{Fig006}. The black curve shows one of 9 bellows and a part of the first from 8 TESLA cavities of cryomodule A1. Each TESLA cavity is composed by 9 cells. The red curve presents the taper at the entrance of the high harmonic cryomodule AH1 which includes again 8 cavities and 9 bellows. The blue curve describes the geometry of the transverse deflecting cavity (TDS) which includes 16 cells. In all cavities the first and the last cells have a little different shapes. The details of the geometry can be found in~\cite{Zag03, TDS1, AH1} and in the papers cited therein.

For an approximate analytical form of wake functions $w_{z}(s)$, $w_{\perp}(s)$ we use modified and extended forms of the expressions suggested in ~\cite{Bane87,  Bane97, Bane03, Zag03}. We present the longitudinal and the transverse wake functions as follows:
\begin{align}\label{EQ_WakeL}
w_z(s)=\theta(s)\bigg(A e^{-\sqrt{k_0 s}}+B\frac{\cos((k_1 s)^{\alpha})}{\sqrt{s}+
	k_2 s}+C \cos(k_3 s)+D\delta(s)\bigg),
\end{align}
\begin{align}\label{EQ_WakeT}
w_{\perp}(s)=\theta(s)\bigg(A'\Big(1-(1+\sqrt{k'_0 s}) e^{-\sqrt{k'_0 s}}\Big)+B'\frac{\sqrt{s}}{1+k'_1 s}+C' \sin(k'_2 s)\bigg),
\end{align}
where $\theta(s)$ is Heaviside step function.
The terms with coefficients $A$, $A'$ have the short-range asymptotics of the wake function of an infinite chain of cavities~\cite{Bane03}. The terms with coefficients $B$, $B'$ have the short range asymptotics of the wake function of an isolated cavity~\cite{Bane87}. The terms with coefficients $C$, $C'$ approximate a long term behaviour of the wake at the tail of the relatively long bunch in the injector section. Finally, the delta-function term with coefficient $D$ is responsible for the contribution of step-out or collimator like geometry features. The first cosinus term in Eq.~(\ref{EQ_WakeL}) is introduced in order to describe an oscillation in the wake potentials. This oscillation appears in a quite short bunch length when the main contribution moves from the first term (periodic model) to the second term (isolated cavity).  For example, for the case of high harmonic cryomodule AH1 we see it in Fig.~\ref{Fig007} for the bunches shorter than 50 $\mu$m.

\begin{table}[htbp]
	\centering
	\caption{The coefficients of the function $w_z(s)$.}
	\label{Table_L}
	\begin{tabular}{lccccccccc}
		\hline\hline
		&\boldmath $A$ &\boldmath $k_0$ &
		\boldmath $B$ &\boldmath $k_1$ &
		\boldmath $\alpha$ &\boldmath $k_2$ &
		\boldmath$C$&\boldmath $k_3$ &\boldmath $D$\\
		&V/pC&	1/mm &V/pC&1/mm&	&1/mm &V/pC&1/mm&V/pC\\
		\hline
		{\bf A1}	&287&	0.413 &4.75&35.0&	0.746&1.52 &12.1&0.181&0\\
		{\bf AH1}	&711&	1.94&1.50&106&	0.661&0.201 &49.2 &0.154&0.0114\\
		{\bf TDS}	&40.7&	0.239 &0.524&6.54&	0.655&0.140 &1.57&0.159&0\\
		\hline\hline
	\end{tabular}
\end{table}

\begin{table}[htbp]
	\centering
	\caption{The coefficients of the function $w_{\perp}(s)$.}
	\label{Table_T}
	\begin{tabular}{lcccccc}
		\hline\hline
		&\boldmath $A'$ &\boldmath $k'_0$ &
		\boldmath $B'$ &\boldmath $k'_1$ &
		\boldmath $C'$ &\boldmath $k'_2$ \\
		&V/pC/mm&	1/mm &V/pC/mm&1/mm&V/pC/mm&1/mm \\
		\hline
		{\bf A1}	&1.03&	0.426 &3.63&0.0664&	0.113&0.232 \\
		{\bf AH1}	&2.75&	2.47&22.9&0.00968&	0.645&0.287 \\
		{\bf TDS}	&0.0863&	0.714 &3.81&0.0135&	0.0675&0.181\\
		\hline\hline
	\end{tabular}
\end{table}

The coefficients of the longitudinal wake function $w_{z}(s)$ are found from numerical simulations with code ECHO~\cite{ECHO} for the Gaussian bunches with an rms length between 2 mm and 10 $\mu$m. The longitudinal wake potentials for the Gaussian bunch profiles $\lambda_{\sigma_i}(s)$
\begin{align}\label{Eq_conv}
W_{z,i}(s)=[w_z*\lambda_{\sigma_i}](s)=\int_{-\infty}^{s}w_z(s-s')\lambda_{\sigma_i}(s')ds', 
\end{align}
are fitted to the numerical results to minimize the residual $\delta_z$ defined as
\begin{align}
\delta_z = \sum_{i}\lVert \hat{W}_{z,i} -W_{z,i} \rVert , 
\end{align}
where {$\lVert \cdot \rVert$  is $L^2$-norm  and} $\hat{W}_{z,i}$ are the wake potentials obtained  numerically with code ECHO for longitudinal bunch shapes $\lambda_{\sigma_i}$, $\sigma_i=$ 10, 15, 25, 50, 250, 500, 100, 2000 $\mu$m. Table~\ref{Table_L} lists the coefficients of Eq.~(\ref{EQ_WakeL}).

The coefficients of the transverse wake function $w_{\perp}(s)$ are found from the fit of the transverse wake potentials 
\begin{align}
W_{\perp,i}(s)=\int_{-\infty}^{s}w_{\perp}(s-s')\lambda_{\sigma_i}(s')ds', 
\end{align}
to the numerical results in the same manner. Table~\ref{Table_T} lists the coefficients of Eq.~(\ref{EQ_WakeT}).

Figure~\ref{Fig007} presents the longitudinal and the transverse wake potentials of the Gaussian bunches in the high harmonic cryomodule AH1. The bunch length is varied between 10 $\mu$m and 2 mm. The numerical results from code ECHO are shown by solid blue curves. The analytical approximations obtained with Eq.~(\ref{EQ_WakeL}) and Eq.~(\ref{EQ_WakeT}) are shown by dashed red curves.

The longitudinal and the transverse wake potentials of different RF modules for the electron bunch with charge of 500 pC (obtained in the simulations at the gun phase offset $\Delta\theta=-2^o$) are shown in Fig.~\ref{Fig008}.

\subsection{Particle tracking with collective effects after RF gun}\label{subsec4.3}

The tracking of particles from position $z=3.2$ meters (entrance of cryomodule A1) to the position of OTR screen is done in OCELOT\cite{OCELOT}. It is done in the same way as, for example, in  Elegant~\cite{Elegant}, in which quadrupoles, dipoles, sextupoles, RF cavities and other lattice elements are modeled by linear and second order maps. The focusing effect of RF cavities is taken into  account according to the Rosenzweig-Serafini model ~\cite{Rosenzweig}.  

The space charge forces are calculated by solving the  Poisson equation in the bunch frame. Then the Lorentz-transformed electromagnetic field is applied as a kick in the laboratory frame. For the solution of the Poisson equation we use an integral representation of the electro-static potential by convolution of the free-space Green's function with the charge distribution. The  convolution equation is solved with the help of the fast Fourier transform. The same algorithm for the solution of the three-dimensional Poisson equation is used, for example, in ASTRA.  

The longitudinal wake function near the reference trajectory is presented through the second order Taylor expansion in the way suggested in~\cite{Dohlus_Wake}. {The coefficients $w_z(s)$ and $w'_{\perp}(s)$  of  Eq.~(\ref{Eq_wz}) are singular at $s=0$ and $w_z(s)$  includes a term with $\delta$-function. In order to tabulate these functions we use the representation~\cite{Zagorodnova}
\begin{equation}\label{Eq_w}
w(s)=w_0(s)+R c \delta(s)+c\frac{\partial}{\partial s} \left(w_1(s)\right),
\end{equation} 
where $w_0, w_1 $ are non-singular functions, which can be tabulated easily and the constant $R$ has the meaning of resistivity. 

The function  $w_z(s)$ defined by Eq.~(\ref{EQ_WakeL}) fits in representation of Eq.~(\ref{Eq_w}) as follows
\begin{align}\label{}
w_0(s)&=\theta(s)\bigg(A e^{-\sqrt{k_0 s}}+B\frac{\cos((k_1 s)^{\alpha})-1}{\sqrt{s}+
	k_2 s}+C \cos(k_3 s)\bigg),\quad w_0(+0)=A+C,\\
w_1(s)&=\theta(s)2 B\frac{\log(1+k_2\sqrt{s})}{c k_2},\quad R=\frac{D}{c}.
\end{align} 

The function $w'_{\perp}(s)$  defined by Eq.~(\ref{Eq_ws}) and Eq.~(\ref{EQ_WakeT})  fits in representation of Eq.~(\ref{Eq_w}) as folows
\begin{align}\label{}
	w_0(s)\equiv 0,\quad w_1(s)=c^{-1}w_{\perp}(s),\quad  R=0.
\end{align}

The wake potential for an arbitrary bunch shape $\lambda(s)$ and the wake function $w(s)$ as in  Eq.~(\ref{Eq_w}) can be found by formula 
\begin{align}
W(s)=[w_0*\lambda](s)+R c \lambda(s) + c[ w_1(s)*\lambda'](s),\nonumber
\end{align} 
where $c$ is the light velocity in vacuum, $\lambda'$ is a derivative of charge density $\lambda$ and the asterisk means the convolution as defined in Eq.~(\ref{Eq_conv}). }

In the experiment we have matched the beam to the design optics for each charge at the gun phase offset $\Delta\theta=-2$. We have reproduced this step in the simulations as well. Note, in addition, that the efforts have been spent also with the particle tracking simulations in code ASTRA. It is worth mentioning that a good agreement of simulation results has been reached between code ASTRA and code OCELOT.

{Finally, let us consider modeling of TDS. The particles have been tracked through TDS in Ocelot coordinates $ \left (y, y' = \frac{p_y}{p_0},  c\Delta t, p = \frac{\Delta E}{p_0 c} \right)$ with the linear matrix
\begin{align}
	\label{EqMopen}
	\begin{pmatrix}
		1&L&-\frac{L K}{2} \cos(\psi)&0\\
		0&1&-K \cos(\psi)&0\\
		0&0&1&\frac{L}{1-\gamma^2}\\
		K \cos(\psi)&\frac{L K}{2} \cos(\psi)&- \frac{L K^2 }{6} \cos(2\psi)&1
	\end{pmatrix},
\end{align}
where $\psi$,  $L$  are  the phase and the length of TDS. The parameter $K$ was defined in Eq.~(\ref{Eq_K}). 

In the experiment we had $K=0.38$ $\text{m}^{-1}$, $L=0.7$ m and the mesurements had been done at the phases of $0^o$ and $180^o$. Hence the energy chirp induced by TDS can be estimated as $dE/dt=0.7$ keV/ps and it is one order of magnitude smaller in comparison with the chirp introduced by the collective effects.}

\section{Comparison between measurements and simulations}\label{sec5}

Finally, we present a direct comparison of the measured curves with those obtain from numerical simulations and clarify the impact of different collective effects on the beam dynamics. 

\begin{figure}[htbp]
	\centering
	\includegraphics*[height=75mm]{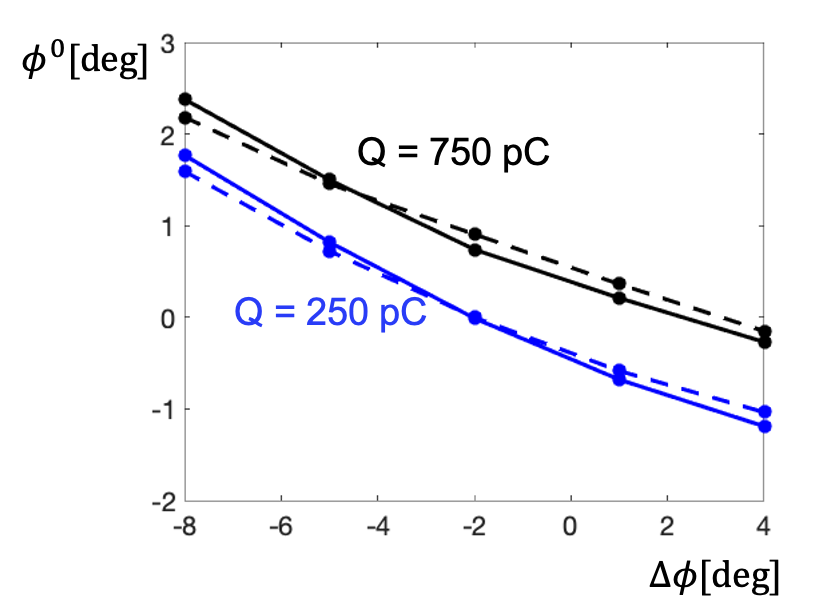}
	\caption{Evolution of the synchronous phase $\hat{\phi}^0$ of cryomodule A1 due to change of the gun phase offset  $\Delta\theta$. The dashed lines present the measured data. {The solid lines are obtained by simulations with collective effects included.}}\label{Fig009}
\end{figure}

In Fig.~\ref{Fig009}, the dots connected by the solid lines represent the synchronous phase $\phi$ of cryomodule A1 obtained from the simulations. The black line shows the change of the synchronous phase $\phi$ in dependence from the gun phase offset $\Delta\theta$ for the bunch charge $Q=750$ pC while the blue line illustrates the case for 250 pC. {The data from the simulations are shown by points connected by dashed lines. The rms error between the measured and  the simulated synchronous phases is equal to 0.1 degree for charge of 250 pC and is equal to 0.15 degree for charge of 750 pC. This means, that the flight time of the bunch in the RF gun obtained in simulations agrees with the experimental data of Table~\ref{Table_A1}. Hence  the gun phase $\theta_{MM}$ is defined correctly in the experiment and in the simulations.}

\begin{figure}[htbp]
	\centering
	\includegraphics*[height=60mm]{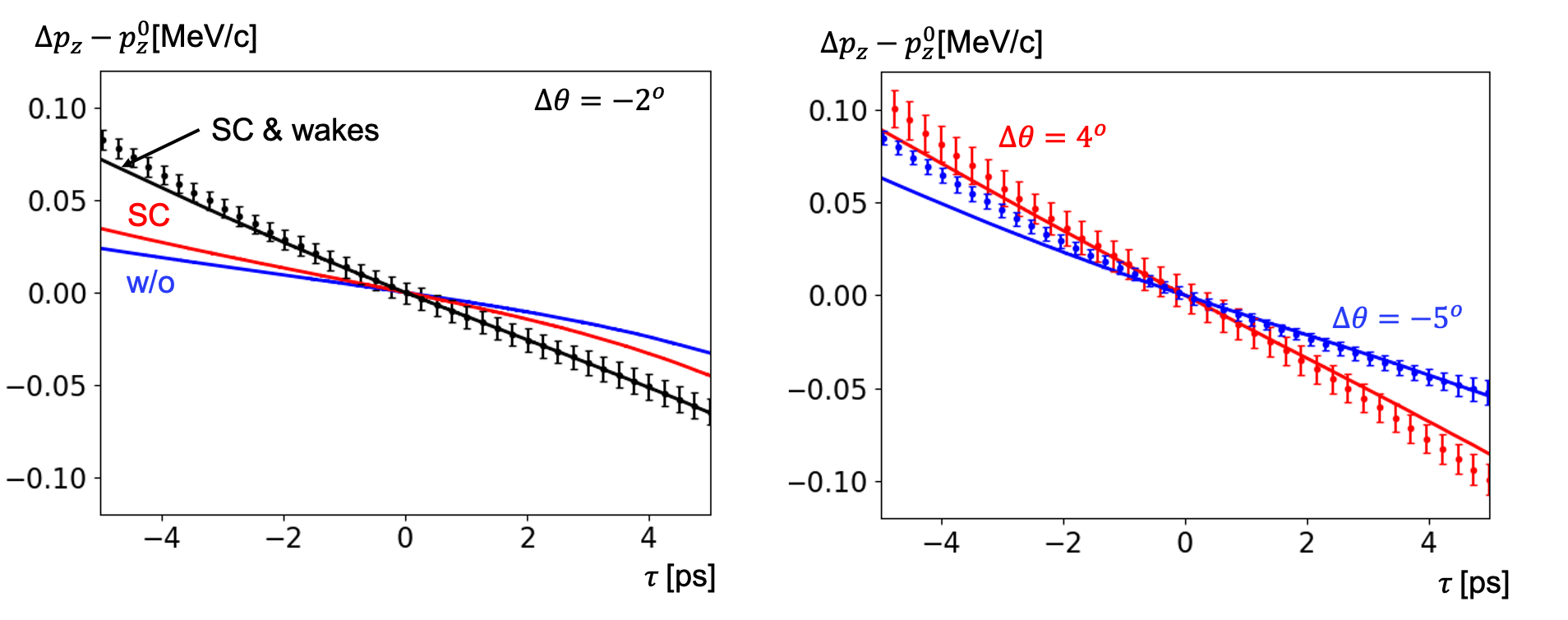}
	\caption{Comparison of the measurements with simulations for the bunch charge $Q=250$ pC. The dots with error bars represent the measurement data. The solid lines show the results obtained from the simulations. {In the right plot the solid lines are obtained by simulations with collective effects included.}}\label{Fig010}
\end{figure}

\begin{figure}[htbp]
	\centering
	\includegraphics*[height=60mm]{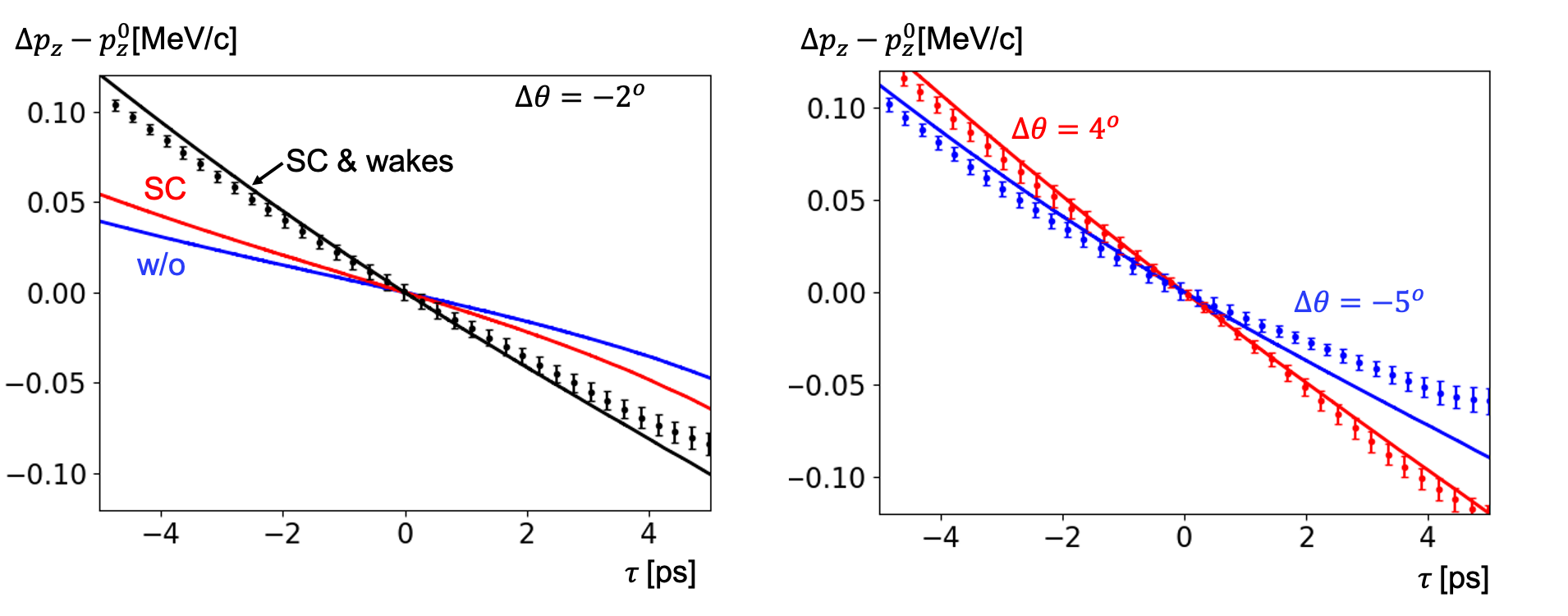}
	\caption{Comparison of the measurements with simulations for the bunch charge $Q=500$ pC. The dots with error bars  represent the experimental results. The solid lines show the results of the simulations. {In the right plot the solid lines are obtained by simulations with collective effects included.}}\label{Fig011}
\end{figure}

\begin{figure}[htbp]
	\centering
	\includegraphics*[height=60mm]{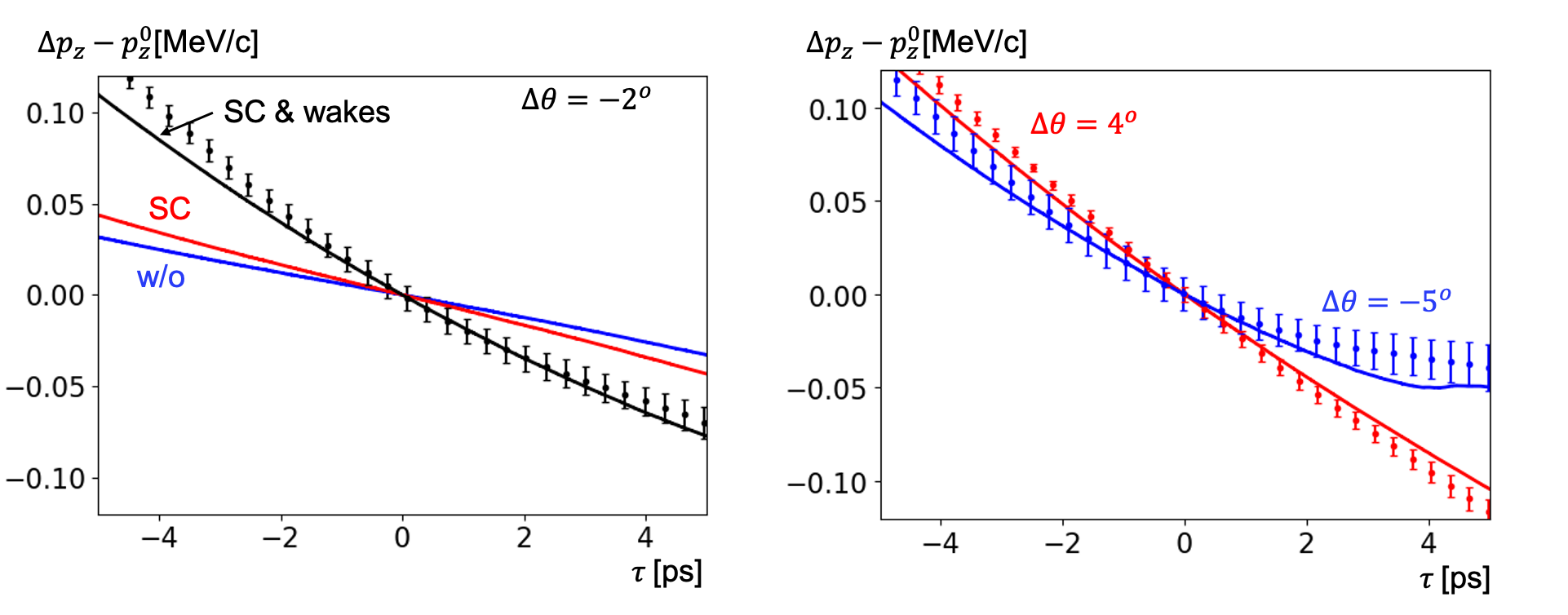}
	\caption{Comparison of the measurements with simulations for the bunch charge $Q=750$ pC. The dots with error bars represent the measurement data. The solid lines show the simulation results. {In the right plot the solid lines are obtained by simulations with collective effects included.}}\label{Fig012}
\end{figure}

The following plots present the residual mean slice momentum $\Delta p_z(t)$ measured in the experiment and obtained by the simulations. We will simply refer it to the mean slice momentum in the following.

Fig.~\ref{Fig010} presents the results for the bunch charge of 250 pC. The dots with error bars on the left plot present the measured mean slice momentum $\Delta p_z(t)$. The error bars show rms error  $\sigma(\tau,\phi^0)$ as defined by Eq.~(\ref{Eq_sigma}). The black solid line presents the mean slice momentum obtained from the simulations as the effects of space charge and the wake fields after the RF gun are taken into account. The red line shows the simulated mean slice momentum if only the space charge effect is considered while the wake fields effects after the RF gun are neglected. Finally, the blue line illustrates the simulation results when all the collective effects after the RF gun are neglected. This curve is defined mainly by the space charge dominated beam dynamics in the RF gun and the chosen gun phase offset $\Delta\theta$. As shown, the left plot presents the results for the gun phase offset $\Delta\theta=-2^o$ while the right plot shows for other gun phases the comparisons between simulations (solid lines) and measurements (the dots with error bars). More specifically, the blue curves represent the results for the gun phase  offset $\Delta\theta=-5^o$ and the red curves present the results for the gun phase offset $\Delta\theta=4^o$. It can be seen that the simulations reproduce the evolution of the energy chirp, correspondingly, with the evolution of the gun phase.

Fig.~\ref{Fig011} and Fig.~\ref{Fig012} present another two sets of comparisons for the cases of 500 pC and 750 pC. The blue solid line in the right plot of Fig.~\ref{Fig012} shows some peculiar shape in the bunch tail. It can be explained by the space charge limited regime~\cite{Chen} for extracting the bunch charge from the cathode at the gun phase offset of $-5^o$. For this special case we were able to extract from the cathode only 700 pC with a laser aperture size of 1.5 mm in both the experiments and the simulations.

Let us define the error $\delta$ between the simulated curve $f(\tau)$ and the measured one $g(\tau)$ on interval $[\tau_1,\tau_2]$
\begin{align}\label{}
	\delta=\frac{\lVert f(\tau)-g(\tau)\rVert}{\sqrt{\tau_2-\tau_1}},
\end{align} 
where $\lVert \cdot \rVert$ is $L^2$ norm. For the gun phase offset  $\Delta\theta=-2^o$  (see left plots in Figs.~\ref{Fig010}-\ref{Fig012}) we summarized this error  in Table~\ref{Table_RMS}. 

\begin{table}[htbp]
	\centering
	\caption{The error $\delta$ in keV between the measured and the simulated  residual mean slice momentum $\Delta p_z(t)$  for the gun phase offset  $\Delta\theta=-2^o$.}
	\label{Table_RMS}
	\begin{tabular}{lccc}
		\hline\hline
		&\bf 250 pC&\bf 500 pC&
		\bf 750 pC \\
		\hline
		{\bf w/o collective effects}	&	26 &33&41\\
		{\bf with space charge only}	&20&	25&35\\
		{\bf with collective effects}	&3&	6 &7\\
		\hline\hline
	\end{tabular}
\end{table}

\begin{figure}[htbp]
	\centering
	\includegraphics*[height=60mm]{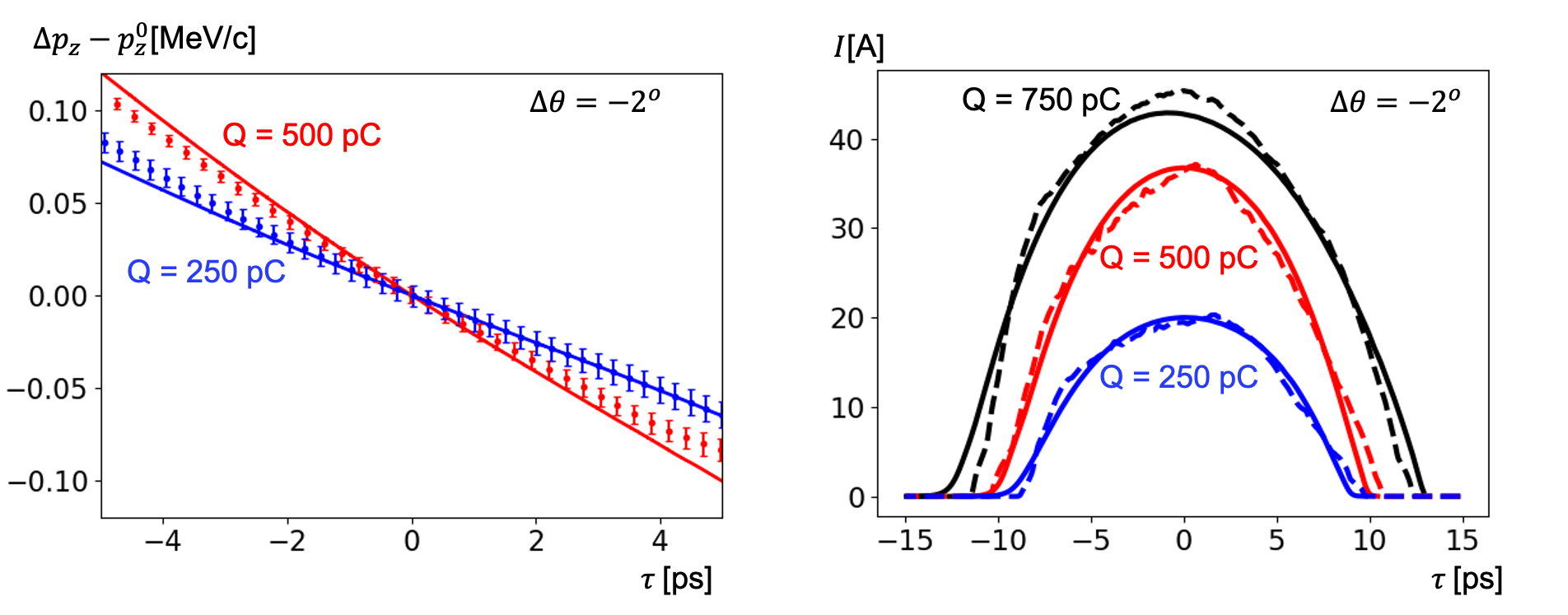}
	\caption{The left plot compare the residual mean slice momentum for different bunch charges at the gun phase offset $\Delta\theta=-2^o$. The right plot compares the current profiles for different bunch charges at the same gun phase. The dashed lines present the results of the measurements. {The solid lines are obtained by simulations with collective effects included.}}\label{Fig013}
\end{figure}

In the left plot of Fig.~\ref{Fig013} we compare the mean slice momentum for bunch charges of 250 pC and 500 pC. As expected, the case with a higher bunch charge shows a stronger energy chirp. The energy chirp for the case of 750 pC is very close to that of the 500 pC case (it is not shown on this plot, see Fig.~\ref{Fig012}). It can be explained by a considerable longer current profile of the bunch, as shown in the right plot of Fig.~\ref{Fig013}. Moreover, for different bunch charges, the right plot of Fig. ~\ref{Fig013} shows the current profiles obtained from both the measurements (dashed lines) and the simulations (solid lines). The difference in the bunch length explains the shift of the synchronous phase as listed in the columns of Table~\ref{Table_A1} for different bunch charges at the same gun phases. It correlates well with the bunch length difference: $\Delta \phi=\omega(\sigma_{Q}-\sigma_{\text{250pC}})$, where $Q$ is equal to 500 pC or 750 pC, and $\sigma_{Q}$ is the bunch length for the bunch charge $Q$.

{From the results presented in Figs.~\ref{Fig010}-\ref{Fig013} and Table~\ref{Table_RMS} we can conclude that the reliability and performance of the physical models applied in the simulations for modeling the space charge forces and the wake fields are confirmed by the systematical measurements. Additionally, we have shown that after the RF gun the wake fields have much stronger impact on the beam dynamics than the space charge forces. The physical models used in the simulation of the beam dynamics after the gun allow to reduce the absolute error in the modeling of the correlated energy chirp by order of magnitude in comparison with the case when the collective effects after the gun  are neglected.}

\section{Summary}\label{sec6}

We have considered both measurements and simulations for the impact of the collective effects on the longitudinal phase space of the electron bunch at the injector section of the European XFEL. The time-resolved measurements of mean slice momentum  along the electron bunch prove the theoretical models of collective effects {used in the simulations}. Moreover, we have suggested a method for accurately identifying the synchronous phase of the RF module. It allows to extract the correlation in the mean slice momentum due to the RF fields and simplifies the comparison between measurement and simulation and the corresponding analysis. Additionally, a new model of the wake function in the finite chain of RF cavities is suggested and cross-checked with the simulations as well as the measurements.   

\section*{Acknowledgments}\label{sec7}

The authors thank W. Decking, M. Dohlus  and M. Krasilnikov for helpful discussions. We thank members of the European XFEL team for providing help and conditions to carry out the measurements.

\end{document}